\documentclass[aps,pre,twocolumn,showpacs,superscriptaddress]{revtex4}
\usepackage{mathrsfs}
\usepackage{graphicx}
\usepackage{bm}
\usepackage{amsmath,amssymb}

\usepackage{color}

\renewcommand{\v}{{\bf v}}
\renewcommand{\r}{{\bf r}}

\newcommand{\beq}{\begin{equation}}
\newcommand{\eeq}{\end{equation}}
\newcommand{\beqa}{\begin{eqnarray}}
\newcommand{\eeqa}{\end{eqnarray}}

\begin{document}
\title{Hydrodynamic simulations of self-phoretic microswimmers}
\author{Mingcheng Yang}
\email{mcyang@iphy.ac.cn} \affiliation{Theoretical Soft-Matter and
Biophysics, Institute of Complex Systems,\\
Forschungszentrum J\"ulich, 52425 J\"ulich,
Germany}\affiliation{Beijing National Laboratory for Condensed
Matter Physics and Key Laboratory of Soft Matter Physics,
\\Institute of Physics, Chinese Academy of Sciences, Beijing 100190,
China}
\author{Adam Wysocki}
\affiliation{Theoretical Soft-Matter and Biophysics,
Institute of Complex Systems,\\
Forschungszentrum J\"ulich, 52425 J\"ulich, Germany}
\author{Marisol Ripoll} \email{m.ripoll@fz-juelich.de}
\affiliation{Theoretical Soft-Matter and Biophysics,
Institute of Complex Systems,\\
Forschungszentrum J\"ulich, 52425 J\"ulich, Germany}
\date{\today}

\begin{abstract}
A mesoscopic hydrodynamic model to simulate synthetic self-propelled
Janus particles which is thermophoretically or diffusiophoretically
driven is here developed. We first propose a model for a passive
colloidal sphere which reproduces the correct rotational dynamics
together with strong phoretic effect. This colloid solution model
employs a multiparticle collision dynamics description of the
solvent, and combines potential interactions with the solvent, with
stick boundary conditions. Asymmetric and specific colloidal surface
is introduced to produce the properties of self-phoretic Janus
particles. A comparative study of Janus and microdimer phoretic
swimmers is performed in terms of their swimming velocities and
induced flow behavior. Self-phoretic microdimers display long range
hydrodynamic interactions and can be characterized as pullers or
pushers. In contrast, Janus particles are characterized by short
range hydrodynamic interactions and behave as neutral swimmers. Our
model nicely mimics those recent experimental realization of the
self-phoretic Janus particles.
\end{abstract}



\maketitle

\section{Introduction}

Synthetic microswimmers have recently stimulated considerable
research interest from
experimental~\cite{paxton,dreyfus,golestanian07,kapral10sma,jiang10,bechinger11}
and theoretical
viewpoints~\cite{golestanian05,golestanian12,seifert12jcp}. This is
due to their potential practical applications in lab-on-a-chip
devices or drug delivery, and fundamental theoretical significance
in non-equilibrium statistical physics and transport processes.
Self-phoretic effects have shown to be an effective and promising
strategy to design such artificial
microswimmers~\cite{anderson,golestanian05,golestanian07b,golestanian07,kapral10sma,jiang10,golestanian14},
where the microswimmers are driven by gradient fields locally
produced by swimmers themselves in the surrounding solvent. In
particular, the collective behavior of a suspension of
self-diffusiophoretic swimmers has recently been studied in
experiments~\cite{bocquet10,bocquet12,chaikin13sci,bechinger13prl}.

Self-phoretic swimmers are typically composed of two parts: a
functional part which modifies the surrounding solvent properties
creating local gradient fields, and a non-functional part which is
exposed then to the local field gradients. Most existing
experimental investigations of the self-phoretic microswimmers
consider Janus particles, which can be quite easily synthesized
using partial metal coating on colloidal
spheres~\cite{golestanian07,jiang10}. In diffusiophoretic
microswimmers, the metal coated part catalyzes a chemical reaction
to induce a concentration gradient. In thermophoretic microswimmers,
the metal coated part is able to effectively absorb heat from e.g.
an external laser, which creates a local temperature gradient. The
investigations performed by computer simulations have mostly
considered dimer structures composed of two connected beads instead
of Janus particles~\cite{kap07b,kapral10sm,yang11,kapral12}. This is
motivated by the simplicity of the structure which can be approached
by a two beads model. Janus particles have been recently simulated
by employing a many beads model~\cite{zerbetto,kapral13nano}, which
has provided an interesting but computationally costly approach. The
fundamental differences on the hydrodynamic behavior of Janus and
dimer swimmers, as well as the interest in the investigation of
collective phenomena of these systems strongly motivates the
development of simple and effective models to simulate the
self-phoretic Janus particles.

A single-bead model of the self-phoretic Janus particle in solution
is here proposed, together with a detailed comparative study of the
hydrodynamic properties of dilute solutions of both self-phoretic
Janus particle and microdimer. While the solvent is explicitly
described by a coarse grained approach known as multiparticle
collision dynamics (MPC), it is necessary to develop a description
of a colloidal particle able to produce strong phoretic effect, and
reproduce the correct rotational dynamics. The proposed colloid
model combines potential interactions with the solvent with stick
hydrodynamic boundary conditions, such that integrate the above two
properties into a single bead. The properties of self-phoretic Janus
particles are introduced then with asymmetric and specific particle
surface. The validity of the model is proved by implementing the
simulations of both the self-diffusiophoretic and
self-thermophoretic microswimmers. The flow field induced by the
self-phoretic Janus particle is measured and compared with that
around the self-phoretic dimer and their analytical predictions. The
efficiency of the model and the consistency of the results puts this
method forward as a reliable and powerful tool to investigate the
collective behavior of self-phoretic microswimmers.

\section{Simulation of a Janus microswimmer in solution}

The typical sizes and time scales of a Janus colloidal particle and
the surrounding solvent particles are separated by several orders of
magnitude which are impossible to cover with a microscopic
description. Over the last decades various mesoscopic simulation
methods have been developed to bridge such an enormous gap. Here, we
employ an especially convenient hybrid scheme that describes the
solvent by MPC which is a coarse-grained particle-based
method~\cite{kap99,kap00,pre05,pad06,kapral08acp,gompper09}, while
the interactions of the Janus particle with the solvent are
simulated by standard molecular dynamics (MD).

MPC consists of alternating streaming and collision steps. In the
streaming step, the solvent particles of mass $m$ move ballistically
for a time $h$. In the collision step, particles are sorted into a
cubic lattice with cells of size $a$, and their velocities relative
to the center-of-mass velocity of each cell are rotated around a
random axis by an angle $\alpha$. In each collision, mass, momentum,
and energy are locally conserved. This allows the algorithm to
properly capture hydrodynamic interactions, thermal fluctuations, to
account for heat transport and to maintain temperature
inhomogeneities~\cite{luesebrink12a,yang14sm}. Simulation units are
chosen to be $m=1, a=1$ and $k_B\overline{T}=1$, where $k_B$ is the
Boltzmann constant and $\overline{T}$ the average system
temperature. Time and velocity are consequently scaled with
$(ma^2/k_B\overline{T})^{1/2}$ and $(k_B\overline{T}/m)^{1/2}$
respectively.  The solvent transport properties are determined by
the MPC parameters~\cite{ihl03c,tuz06}. Here, we employ the standard
MPC parameters $\alpha=120^\circ$, $h=0.1$, and the mean number of
solvent particles per cell $\rho=10$, which corresponds to a solvent
with a Schmidt number $Sc=13$.  The simulation system is a cubic box
of size $L=30a$ with periodic boundary conditions.

By construction, a Janus particle has a well-defined orientation
with a corresponding well-defined rotation, and surface properties
are different in the two colloid hemispheres.  In previous studies
of colloid phoresis with
MPC~\cite{kap07,kapral10sm,yang11,luesebrink12b,yang13sm}, a central
type of interaction such as the Lennard Jones potential has been
employed, which does not result in a rotational motion. Other
studies of rotational colloidal dynamics~\cite{pad05} have employed
MPC with boundary conditions that are in fact thermalized stick
conditions since a surface colloidal temperature needs to be
imposed. In this work, we first modify existing techniques to
construct a specific model that allows us to simulate a colloid with
stick boundary conditions together with potential interactions with
the solvent that locally conserve not only mass and momentum, but
also energy. Then, in order to reproduce the properties of a Janus
particle, the spherical colloid is divided in two hemispheres
characterized by different interactions with the surrounding
solvent. One of this halves (with a polar angle $\theta\leq\pi/2$
with respect to a defined colloid axis ${\bf n}$) is considered to
be the {\em functional} part, while the other half is the {\em
non-functional} part. The functional part of the Janus particle is
where the material has special properties like enabling a chemical
reaction (catalytic) or carrying a high temperature due to a larger
heat adsorption. The special behavior of the functional part
originates local gradients (as of concentration or temperature)
which will induce a phoretic force applied to one or both halves of
the Janus colloid.  In the following sections we introduce first the
model for a colloid with stick boundary conditions and a
well-defined orientation, and then consecutively the thermophoretic
and diffusiophoretic Janus particles.

\subsection{Passive colloid with stick boundary: simulation model}

\begin{figure}[h]
\centering
\begin{minipage}[c]{0.5\textwidth}
\centering
\includegraphics[angle=0,width=3.0in]{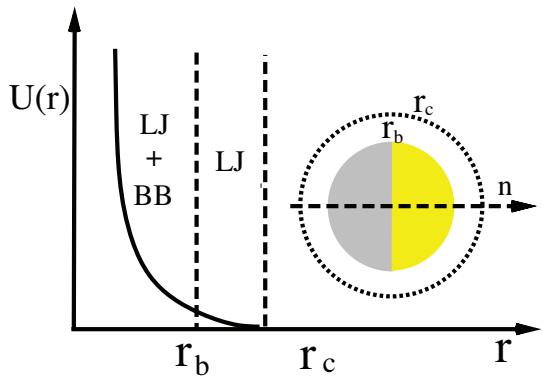}
\end{minipage}
\caption{Schematic diagram of the three regions of the colloid-solvent
  interactions. The inset is a sketch of a Janus
  particle.}\label{fig1}
\end{figure}

A colloidal particle with stick boundary conditions will vary its
direction of motion randomly. This is caused by the stochastic
torque exerted on the particle due to collisions with the solvent.
On a coarse grained level, stick boundary conditions can be modeled
by the bounce back (BB) collision rule~\cite{chen98,lam01}, this is
by reversing the direction of motion of the solvent particle with
respect to the colloidal surface. 
However, the bounce back rule does not induce significant phoretic
effects, such that it is necessary to combine it with a soft
potential.  Practically, we realize this by defining three interaction
regions, as shown in Fig.~\ref{fig1}, where $r$ is the distance
between a solvent particle and the center of a colloid. For distances
larger than the cutoff radius, $r>r_c$, there is no interaction.  For
$r_b>r>r_c$ just the soft central potential is considered. And for
$r<r_b$, both the soft potential and the bounce back collision are
taken into account. The value of the {\em bounce-back radius} $r_{b}$
should be large enough to ensure that a certain amount of solvent
particles participate in the bounce-back collision such that a
significant rotational friction is induced.  On the other hand, the
value of $r_{b}$ should also be small enough such that the
colloid-solvent potential effectively contribute to the phoretic
force.

The interaction potential employed in this work is of
Lennard-Jones~(LJ) type~\cite{vliegent99}, with the general form
\begin{equation}
U(r)= 4\epsilon\left[\left(\frac{\sigma}{r}\right)^{2k}
-\left(\frac{\sigma}{r}\right)^{k}\right]+C, \ r\leq r_c. \label{lj}
\end{equation}
The positive integer $k$ controls the stiffness of the potential,
and $r_c$ is the potential cutoff radius. The potential intensity is
chosen as one of the system units $\epsilon = k_B\overline{T} = 1$,
and the interaction length parameter as \mbox{$\sigma=2.5a$}.  In
this work we choose $r_b=\sigma$ which is also a good estimation for
the colloid radius. Attractive interactions are obtained with $C=0$,
and $r_c=2.5 \sigma$ and repulsive with $C=\epsilon$ and
$r_c=2^{1/k}\sigma$. The mass of the colloidal particle is set to
$M={4\pi \sigma^3m\rho}/{3}=650m$, such that the colloid is
neutrally buoyant. Between two MPC collision steps, $N_{md}$
molecular dynamics steps are employed. The equations of motion are
integrated by the velocity-Verlet algorithm with a time step $\Delta
t=h/N_{md}$, where we use $N_{md}=50$.

Mostly bounce-back collision considers the interaction between solvent
particles and immobile planar walls, where the particle velocity is
simply reversed. Here in contrast, an elastic collision is performed
when a point-like solvent particle with velocity ${\bf v}$ is moving
towards the spherical colloid and is closer to it than $r_b$, this is
$r<r_b$.  The colloidal particle has a linear velocity ${\bf V}$, an
angular velocity $\omega$, and a moment of inertia $I=\chi M
\sigma^2$, with $\chi=2/5$ the gyration ratio. Since the collision is
now performed with a moving object, the relevant quantity for
the collision is $\widetilde{\bf v}$, namely, the solvent
particle velocity relative to the colloid at the colliding point,
\begin{equation}\label{relativ.vel}
\widetilde{\bf v}={\bf v}-{\bf V}-\omega\times{\bf s},
\end{equation}
where ${\bf s} = {\bf r}-{\bf R}$, with ${\bf r}$ and ${\bf R}$, the
position of the solvent particle and of the center of the colloid,
respectively. In the following, we refer to ${\bf s}$ as the contact
vector and $\widetilde{\bf v}$ as the contact velocity. The
conservation of linear and angular momentum imposes the following
explicit expressions for the post-collision velocities
\begin{align}\label{bounce}
\v'=&\v-{\bf p}/m,\nonumber\\ {\bf V}^\prime=
&{\bf V}+{\bf p}/M,\\  \omega'= &
\omega+\left({\bf s}\times {\bf p}\right)/I.\nonumber
\end{align}
The precise form of the momentum exchange ${\bf p}$ can be
calculated in terms of the normal and tangential components of the
contact velocity $\widetilde{\bf v}_n=\hat{\bf s} \left(\hat{\bf
s}\cdot\widetilde{\bf v}\right)$, and $\widetilde{\bf
v}_t=\widetilde{\bf v} - \widetilde{\bf v}_n$, with $\hat{\bf
s}={\bf s}/|{\bf s}|$ the unit contact vector. Imposing the
conservation of kinetic energy and stick boundary condition (see
calculation details in the Appendix \ref{appendix}) leads to
$\widetilde{\bf v}_n^\prime=-\widetilde{\bf v}_n$ and
$\widetilde{\bf
  v}_t^\prime=-\widetilde{\bf v}_t$, which determines
\begin{equation}\label{momentum}
{\bf p}={\bf p}_n+{\bf p}_t
=2\mu\widetilde{\v}_n + \frac{2\mu\chi M}{\chi M +\mu}{\widetilde\v}_t,
\end{equation}
where $\mu=mM/(m+M)$ is the reduced mass.  This collision rule is
similar to the one used in rough hard sphere
systems~\cite{allen,berne77}, although in the present case the
colliding pair is composed of a point particle and a rough hard
sphere~\cite{whitmer10}. This collision method does not change the
positions of the particles and, consequently, the potential energy
does not vary discontinuously.

\subsection{Passive colloid with stick boundary: simulation results}

\begin{figure}[h]
\centering
\begin{minipage}[c]{0.5\textwidth}
\centering
\includegraphics[angle=-90,width=2.8in]{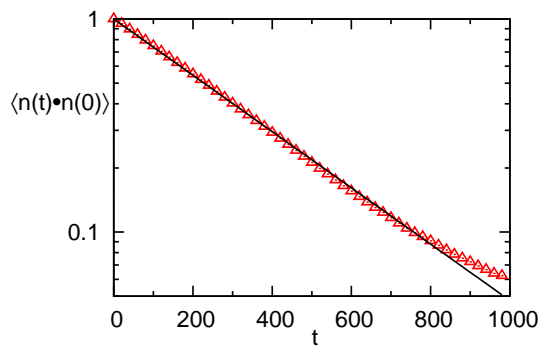}
\end{minipage}
\caption{Time auto-correlation function of the orientation
vector of passive colloidal sphere. Symbols refer to simulation
results, and the line to Eq.~(\ref{dr}).}\label{nnc}
\end{figure}

In order to test the correct rotational dynamics of the proposed
model, we first verify the exponential decay of the
orientational time-correlation function. This is expected to
be~\cite{doi},
\begin{equation}
\langle{\bf n}(t)\cdot{\bf n}(0)\rangle=\exp(-2D_{r}t), \label{dr}
\end{equation}
with the body-fixed orientation vector ${\bf n}$, and $D_{r}$ the
rotational diffusion constant. A repulsive potential with $k=24$ in
Eq.~(\ref{lj}) is chosen for the colloid-solvent interactions. A fit
of Eq.~(\ref{dr}) to our data (shown in Fig.~\ref{nnc}) yields
$D_r=0.0015$ in units of $(k_B\overline{T}/ma^2)^{1/2}$. In order to
provide an analytical estimation of this coefficient, it should be
taken into account that within the cut-off-radius, the number
density of the solvent particles obeys $\rho(r)=\rho
e^{-U(r)/k_{B}T}$ due to the ideal gas equation of state of the MPC
solvent. This results into a position-dependent viscosity. In the
following, we refer to the local number density at the colloid
surface as $\rho_\sigma=\rho(\sigma)=\rho e^{-1}$. The corresponding
dynamic and kinematic viscosity at the particle surface are obtained
using the dependence of $\eta$ on $\rho$ from the kinetic
theory~\cite{tuz06}.  For the MPC solvent employed parameters, we
have that $\eta=7.93$, $\eta_\sigma=2.47$ and
$\nu_\sigma=0.67$. The Stokes-Einstein equation for the rotational
diffusion provides the dependence $D_r=k_{B}T/\zeta_H$, with the
hydrodynamic rotational friction $\zeta_H=8\pi\eta_\sigma \sigma^3$.
With this approximation, we obtain $D_r=0.001$, which
underestimates, but it is still consistent with the simulation
result.

The rotation dynamics can be further analyzed by measuring the angular
velocity autocorrelation function of the colloidal particle. For short
times, Enskog kinetic theory~\cite{davis,pad05} predicts that the
autocorrelation function follows a exponential decay,
\begin{equation}
\lim_{t\rightarrow0}\langle\omega(t)\cdot\omega(0)\rangle=
\langle\omega^2\rangle\exp(-\zeta_{E} t/I),\label{short}
\end{equation}
with $\langle\omega^2\rangle=3k_{B}T/I$, as obtained from energy equipartition
theorem, and $\zeta_{E}$ the Enskog rotational friction coefficient of a sphere suspended in
bath of point-like particles~\cite{whitmer10},
\begin{equation}
\zeta_{E}=\frac{8}{3}
\sqrt{2\pi k_BT\mu}\rho_\sigma\sigma^4\frac{\chi M}{\mu+\chi M}.\label{engkog}
\end{equation}
For long times, the relaxation
of the correlation function is predicted by hydrodynamic
mode-coupling theory~\cite{leeuwen70,pad05} to decay algebraically,
\begin{equation}
\lim_{t\rightarrow\infty}\langle\omega(t)\cdot\omega(0)\rangle=
\frac{3\pi k_BT}{m\rho_\sigma\left(4\pi\nu_\sigma
t\right)^{5/2}}.\label{long}
\end{equation}

\begin{figure}[h]
\centering
\begin{minipage}[c]{0.5\textwidth}
\centering
\includegraphics[angle=-90,width=2.8in]{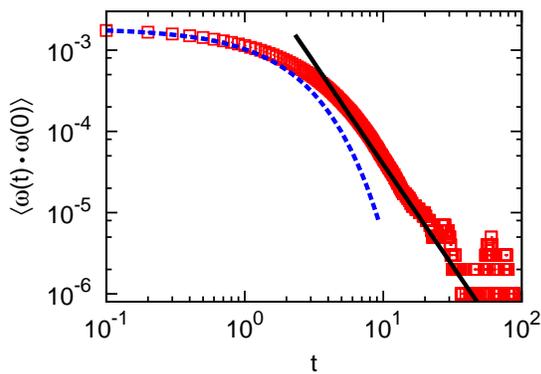}
\end{minipage}
\caption{Time decay of the angular velocity autocorrelation
function of passive colloidal sphere. Symbols correspond to
simulation results, the dashed line to short-time Enskog prediction
in Eq.~(\ref{short}) and solid line to the long-time hydrodynamic
prediction in Eq.~(\ref{long}).}\label{ooc}
\end{figure}

The angular velocity autocorrelation function obtained from
simulations is displayed in Fig.~\ref{ooc} agrees very well with the
theoretical predictions at short and long time regimes respectively
in Eq.~(\ref{short}) and (\ref{long}), where no adjustable parameter
is employed. On the other hand, the rotational diffusion coefficient
can be understood to be determined by the total friction $\zeta$,
with $1/\zeta=1/\zeta_H+1/\zeta_E$. Considering both terms, the
analytical prediction is $D_r=k_BT/\zeta=0.002$, which
overestimates then the measured value of $D_r$. More rigorous
expression for the hydrodynamic rotational diffusion coefficient can
in principle be obtained by solving the Stokes equation with an
inhomogeneous viscosity profile~\cite{kroy12}.

In conclusion, these results ensure that the coarse-graining model
introduced here describes physically correct rotational dynamics
where no surface thermalization has been employed. This is the basic
colloid model on which the Janus structure can be further
introduced.

\subsection{Self-thermophoretic Janus colloid}

A Janus particle partially made/coated with a material of high heat
absorption and heated, for example with a laser, develops around it
an inhomogeneous temperature distribution~\cite{jiang10}. The part
of the Janus particle with lower heat absorption is therefore
exposed to a solvent with a temperature gradient, which originates a
thermophoretic force. Depending on the nature
(thermophilic/thermophobic) of the colloid-solvent interactions in
the non-functional part of the colloid, the thrust will be exerted
towards or against the temperature gradient.

The simulation model combines now the rotating colloid introduced in
the preceding section, with elements of the previously investigated
self-thermophoretic dimer~\cite{yang11}. In particular, the
temperature around the heated hemisphere is fixed by rescaling the
thermal energy of the solvent particles closer than $1.08\sigma$ to
the center of the sphere to a value $T_h$. This means that only a
small layer ($\simeq 0.08\sigma$) around the heated part of the
Janus particle is affected by the rescaling. In this work we have
restricted ourselves to $T_h=1.25\overline{T}$, although a large
range of possible values is accessible.  The inserted energy is
drained from the system by thermalizing the mean temperature of the
system to a fixed value $\overline{T}$. In experiments, the
thermalization is performed at the system boundaries. Although these
two thermalizations are intrinsically different, the differences are
expected to be negligible, when the system is large enough, and
especially when considering the neighborhood of the Janus particle.

\begin{figure}[h]
\includegraphics[width=2.5in]{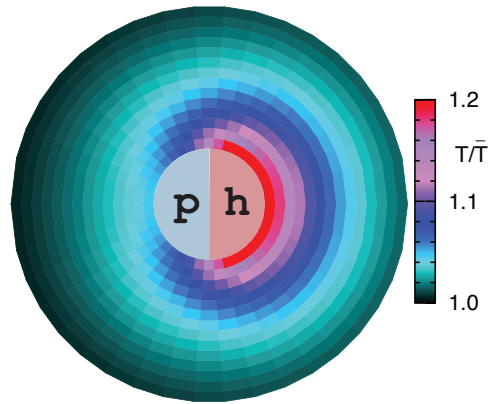}
\caption{Temperature distribution induced by a self-thermophoretic
Janus particle. Here, the Janus particle has a repulsive LJ
potential with $k=3$. The right (h) and the left (p) hemisphere
correspond to the heated and the phoretic parts, respectively.
Because of axis-symmetry, only the distribution in a section across
the axis is displayed.}\label{tm}
\end{figure}

\begin{figure}[h]
\centering
\begin{minipage}[c]{0.5\textwidth}
\centering
\includegraphics[angle=-90,width=2.8in]{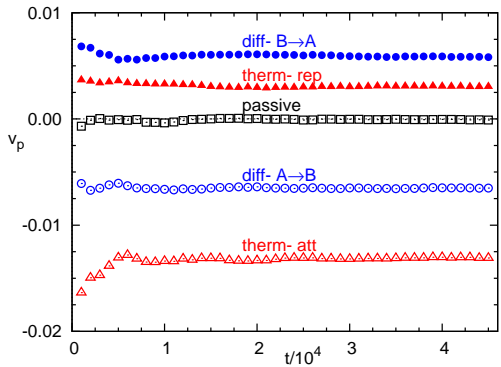}
\end{minipage}
\caption{Self-propelled velocity as a function of time as averaging
parameter. Triangles refer to self-thermophoretic Janus particles
with repulsive and attractive LJ-type potentials. Circles refer
to self-diffusiophoretic Janus particles. Solid symbols refer
to forward motion, namely along the polar axis towards the functional part.
Open symbols refer to backwards motion. For reference, squares denote the
velocity measured for a purely passive colloid. }\label{vp}
\end{figure}

\begin{figure}[h]
\centering
\begin{minipage}[c]{0.5\textwidth}
\centering
\includegraphics[angle=-90,width=2.6in]{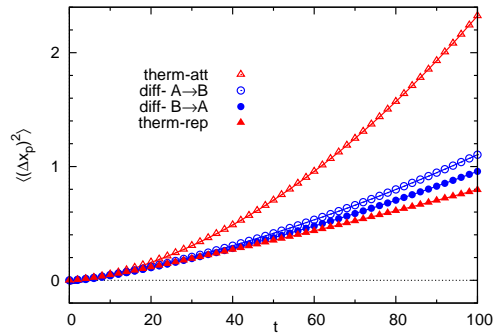}
\end{minipage}
\caption{MSD of the center-of-mass of the Janus particle along the
polar axis as a function of time. Simulation parameters are those of
Fig.~\ref{vp}. Lines correspond to  fits with
Eq.~(\ref{vp2}).}\label{sd}
\end{figure}

Two different colloid-solvent potentials of LJ-type Eq.~(\ref{lj})
are employed in the simulations provided here, a soft
repulsive potential with $k=3$, and a short-range attractive
potential with $k=24$. The particular shape of the colloid-solvent
potential has already shown~\cite{yang11,luesebrink12b} to influence
the magnitude of the thermophoretic force, and interestingly also
its direction. The repulsive LJ potential is expected to produce a
thrust pointing to the heated hemisphere; while the attractive
potential will lead to a driving force in the opposite direction.
Two procedures are employed to quantify the self-propelled velocity
$v_{p}$. A direct characterization can be performed by projecting
the center-of-mass velocity of the Janus particle on its polar axis,
$v_p=\langle{\bf V}\cdot{\bf n}\rangle$. Figure~\ref{vp} shows how
direct measurements of $v_p$ are well-defined for different
interaction potentials as a function of time, which is employed as
an averaging parameter. Indirect determination of $v_{p}$ is
obtained by measuring the mean square displacement (MSD) of the
Janus particle along its polar axis. In this direction, the motion
of the Janus particle can be divided into a pure diffusion and a
pure drift, and it is related to the self-propelled velocity via
\begin{equation}
\langle(\Delta x_{p})^2\rangle=2D_p t+v^{2}_{p} t^{2}.\label{vp2}
\end{equation}
Here $D_p$ is the translational diffusion coefficient of the Janus
particle along its axis. The mean square displacement in the polar
direction is shown in Fig.~\ref{sd} as a function of time for Janus
particles with different colloid-solvent interactions. At very small
times, an initial inertial regime with a quadratic time dependence
is observed.  For times larger than the Brownian time, the diffusive
behavior coexists with the presence of the self-propelled velocity
as indicated in Eq.~(\ref{vp2}). A fit to the data allows us to
determine both $D_p$ and $v_p$ with good accuracy. Direct and
indirect determination of $v_p$ agree very well within the
statistical accuracy, as can be seen in Table~\ref{tab:vp}.

\begin{table}[h]
\begin{tabular}{r|r|r|r|r|}
&therm-att&therm-rep&diff-$A\to B$&diff-$B\to A$\\ \hline
$v_p$ (direct)&-0.0131&0.0030&-0.0065&0.0059\\ \hline
$v_p$ (indirect)&-0.0133&0.0035&-0.0070&0.0059\\ \hline
$D_p$ (indirect)&0.0029&0.0035&0.0032&0.0032\\ \hline\hline
$v_p$ (dimer)&-0.0068&0.0047&& \\ \hline
$D_p$ (dimer)&0.0028&0.0034&& \\ \hline
\end{tabular}
\caption{Summary of the self-propelled velocities, and the diffusion
coefficient of the thermophoretic and diffusiophoretic Janus
particles obtained from the simulations with the direct and indirect
methods. For comparison the same quantities are displayed for
our prior results on the thermophoretic microdimers~\cite{yang11}.
\label{tab:vp}}
\end{table}

The quantitative values of the propelled velocities are determined
by the nature of the thermophoretic forces. As in the case of
thermophoretic microdimers~\cite{yang11}, these forces are related
to the temperature gradients $\nabla T$, and the thermal diffusion
factor $\alpha_T$ which characterizes the particularities of the
colloid-solvent
interactions~\cite{piazza08,wuerger10,luesebrink12b}. The
self-propelled velocity is then $v_p=-\alpha_T\nabla k_B
T/\gamma_p$, with $\gamma_p$ the particle translational frictional
coefficient and $D_p=k_BT/\gamma_p$. The hydrodynamic translational
frictional coefficient is $\gamma^H_p= B \eta \sigma$ with $B$ being
a numerical factor given by the boundary conditions. Colloids with
stick boundary conditions have $B=6\pi$, while colloids with slip
boundary conditions have $B=4\pi$. The here proposed model provides
stick boundary conditions for colloids at $r\simeq r_b$ with the
surface viscosity $\eta_\sigma$, and slip for $r>r_b$, which means
that the overall colloid behavior will be effectively {\em partial
slip}. The stick boundary approach predicts
$k_BT/(6\pi\eta\sigma)\simeq0.0027$, such that the slightly larger
simulation results in Table~\ref{tab:vp} are consistent with the
partial slip prediction. In principle these values should still be
corrected by considering the Enskog contribution and finite size
effects. However, the precise form and validity of these corrections
is still under debate for colloids simulated with
MPC~\cite{pad06,whitmer10}. It can be observed that the values for
the thermophoretic attractive potential are smaller than those for
the repulsive one, which reflects the larger viscosity $\eta_\sigma$
provided by the attractive surface interactions. Interestingly, the
values for $D_p$ of the thermophoretic dimers are very similar than
those of the Janus particles. This can be understood as the result
of two canceling effects. On the one hand the microdimer has larger
size than the Janus particle, which decreases the translational
diffusion. On the other hand, the microdimer is here simulated with
slip boundary conditions, which reduces the friction in comparison
with the stick, or partial slip boundary conditions employed for the
Janus particle. Given that $D_p$ is not significantly changing for
the results in Table~\ref{tab:vp}, the variation of numerical values
of $v_p$ can be related to the differences in $\nabla T$ and
$\alpha_T$. The actual value of $\nabla T$ varies along the particle
surface, and it is not the same for both particle geometries. The
determination of $\alpha_T$ is given by the size, the geometry, and
the specific interactions between the colloid and the solvent. The
comparison of the measured $v_p$ for the dimer and the Janus
particles is therefore non-trivial and deserves a more in-depth
investigation. Furthermore, the bounce-back surface considered in
the Janus particle model produces an additional thermophobic thrust,
which could explain the enhanced value of the Janus particle with
attractive interactions.

In the presence of a temperature gradient, the transport of heat is
a relevant process which in experimental systems occurs in a much
faster time scale than the particle
thermophoresis~\cite{wieg04,jiang10,kroy10}. For thermal energy
propagation the characteristic time is $\tau_\kappa\sim a^2/\kappa$
with $\kappa$ the thermal diffusivity, and the time scale of
particle motion is related to the self-propelled velocity by
$\tau_m\sim a/v_{p}$. Using $\kappa$ estimated from kinetic
theory~\cite{tuz06} and the measured $v_{p}$, we have
$\tau_{\kappa}/\tau_{m}\sim10^{-1}$ for our simulation parameters.
This means that both times are also well-separated in the
simulations, and that temperature profile around the swimmer is
almost time-independent.

\subsection{Self-diffusiophoretic Janus colloid}

A colloidal particle with a well-defined part of its surface with
catalytic properties can display self-propelled
motion~\cite{paxton,golestanian07,bocquet10,bocquet12,sen07prl}.
Such functional or catalytic part of the Janus particle catalyzes a
chemical reaction, which creates a surrounding concentration
gradient of the solvent components involved in the reaction, which
typically have different interactions with the colloid. This
gradient in turn induces a mechanical driving force
(diffusiophoretic force) on the Janus particle and hence propulsion.
The direction of the self-propelled motion will be related to the
interaction of each solvent component with the colloid. Chemical
reactions are generally accompanied by an adsorption or emission of
energy. A catalytic Janus particle could therefore generate a local
temperature gradient which would induce an additional thermophoretic
thrust. However, existing experiments of Pt-Au
micro-rods~\cite{paxton} have shown the contribution of this effect
to be negligible.

\begin{figure}[h]
\includegraphics[width=3.0in]{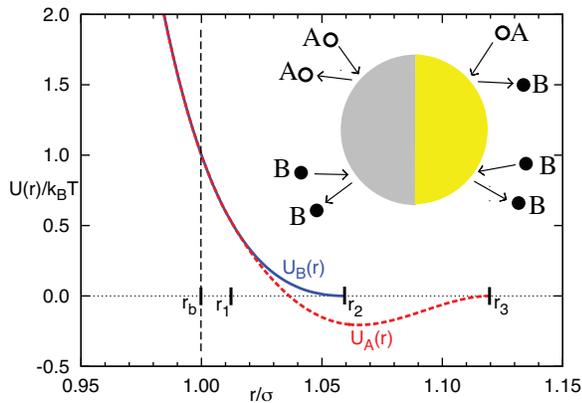}
\caption{Potential interactions between the two solvent species and
the Janus particle. $U_{A}(r)$ is defined in Eq.~(\ref{ua}) and
$U_{B}(r)$ in Eq.~(\ref{lj}). Inset: Schematic representation of the
catalytic and non-catalytic hemispheres of the Janus particle
and the interaction of the A and B species with each
hemisphere.}\label{force}
\end{figure}

The effect of irreversible chemical reactions has already been
included in a MPC simulation study of chemically powered nanodimers
by R\"{u}ckner and Kapral~\cite{kap07b}. Similar to that work, we
here consider a solvent with two species A and B, together with the
model of the stick boundary colloid previously introduced. The
reaction $A \to B$ is performed with a probability $p_R$ whenever an
A-solvent particle is closer than a distance $r_1$ to the catalytic
hemisphere of the Janus particle (see inset of Fig.~\ref{force}).
Besides this reaction and the MPC collision, there are no further
interactions between A and B solvent particles. Another important
element to induce self-propelled motion is that the interaction of
each component with the colloid surface should be
different~\cite{kap07b}. We therefore consider that solvent species
A and B interact with the Janus particle with different potentials
$U_{A}(r)$ and $U_{B}(r)$, but with the same bounce-back rule. A
change of potential energy at the point where the $A \to B$ reaction
occurs could be numerically unstable, and would lead to a local
heating or cooling of the surrounding solvent. In order to model
here a purely diffusiophoretic swimmer, we choose smoothly varying
potentials $U_{A}(r)$ and $U_{B}(r)$ which completely overlap for
$r\leq r_1$, ensuring a reaction without an energy jump. We consider
$U_{B}(r)$ as the repulsive LJ-type potential in Eq.~(\ref{lj}) with
$k=12$. $U_{A}(r)$ in Fig.~\ref{force} is constructed in four
intervals by a cubic spline interpolation, which yields to
\begin{equation}\label{ua}
U_A(r)= \left\{
\begin{array} {ll}
U_B(r)&(r\leq r_{1})\\
a_0+a_1r+a_2r^2+a_3r^3
&(r_{1}\leq r\leq r_2)\\
b_0+b_1r+b_2r^2+b_3r^3
&(r_2\leq r\leq r_3)\\
0 &(r_3\leq r)
\end{array}
\right.
\end{equation}
where the coefficients and the distances to determine the related
intervals are specified in the Table~\ref{tab.ua}.
\begin{table}[h]
\begin{tabular}{|l|l|l|l|}
\hline
$a_0=844.6$&$a_1=-849.7$&$a_2=280.7$&$a_3=-30.3$ \\
\hline
$b_0=3283$&$b_1=-3610$&$b_2=1322$&$b_3=-161.4$ \\
\hline
$r_b= \sigma$
&$r_1= 1.0132\sigma$&
$r_2= 1.06\sigma$&
$r_3=1.12\sigma$\\
\hline
\end{tabular}
\caption{Coefficients employed in the simulations
for the potential function $U_A$ in Eq.~(\ref{ua}).
\label{tab.ua}}
\end{table}

\begin{figure}[h]
\includegraphics[angle=-90,width=3.in]{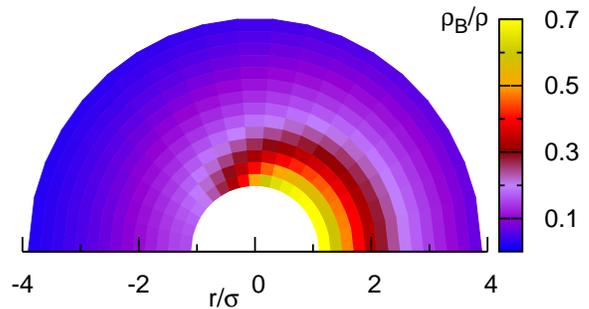}
\caption{Local number density distribution of B-type particles
induced by a self-diffusiophoretic Janus particle with the
\mbox{$A\to B$} reaction. The right hemisphere corresponds to the
catalytic part.}\label{cm}
\end{figure}

Simulations are initiated with a solvent composed only of A-type
particles. The considered chemical reaction $A \to B$ in the
catalytic part of the Janus particle is irreversible, such that
A-type solvent particles are gradually consumed.  In order to keep a
stationary concentration gradient, A-particles are constantly fed
into the system. Concretely, we fix the reaction probability to
$p_{R}=0.1$, and whenever a B-type particle is at a distance $d$
from the Janus particle (we consider $d=5\sigma$), it automatically
converts into A. This allows the system to reach an steady-state
concentration distribution of B molecules around the swimmer.
Figure~\ref{cm} shows $\rho_B$, the number of B-type particles per
unit cell. It can be seen that on the catalytic hemisphere there are
mostly B-type particles, while in the phoretic hemisphere the
situation is reversed and there are mostly A-particles. The
self-propelled velocity is quantified by using the direct and
indirect methods as already described for the self-thermophoretic
Janus particles. The results are displayed separately in
Fig.~\ref{vp} and Fig.~\ref{sd}, and the numerical values are
summarized in Table~\ref{tab:vp} where the nice agreement between
the methods can be observed. The diffusion coefficients for the both
diffusiophoretic Janus particles are the same, which is related to
the fact that at the surface both potentials are the same. The value
of the self-propulsion velocity, $v_p$, is determined by the choice
of the colloid-solvent potentials, the reaction probability and the
boundary conditions. For the considered $A\to B$ reaction with
$U_{B}(r)$ repulsive, and $U_{A}(r)$ attractive, the concentration
gradient pushes the Janus particle against the direction of the
polar axis ${\bf n}$. A reciprocal choice of potentials, which is
almost equivalent to consider the reaction $B\to A$, pushes the
Janus particle along ${\bf n}$ as can be verified in Fig.~\ref{vp}
and Table~\ref{tab:vp}. It should be noted that the values of the
velocities in both simulations are not exactly reversed, since the
reciprocal choice of potentials does not correspond to a perfectly
reverse distribution of the species concentrations. A comparison of
the velocities for the diffusiophoretic Janus particle in this work,
and the existing data for microdimers and Janus
particles~\cite{kap07b,kapral13nano} is not really straightforward
since the employed parameters and potentials are different. The
systems are though not so different, and the values of $v_p$ range
from similar values to approximately four times smaller.

The time scale of the particle motion of a self-diffusiophoretic
swimmer $\tau_m$ needs to be compared with the time scale of solvent molecule
diffusion $\tau_s$, which is a much faster process in experimental systems.
The solvent diffusion coefficient $D_{s}$
determines $\tau_s=a^2/D_{s}$. For the employed simulation parameters,
$D_{s}$ from the kinetic theory, and the measured
$v_{p}$ determine the separation of both time scales to be
$\tau_{s}/\tau_m\sim10^{-1}$.

\section{Flow field around phoretic swimmers}

In the previous section, an efficient model to simulate the behavior
of self-phoretic Janus particles has been introduced, and the
obtained velocities have been related with the employed system
parameters. Another fundamental aspect in the investigation of
microswimmers is the effect of hydrodynamic
interactions~\cite{lauga09}, and how do these compare with the
effect of concentration or temperature gradients.  In the case of
self-propelled particles, the temperature or concentration
distributions decay with $1/r$ around the particle, such that their
gradients decay as $1/r^2$. Furthermore, the hydrodynamic
interactions have shown to be fundamentally different for swimmers
of various geometries and propulsion mechanisms, yielding to
phenomenologically different behaviors classified in three types,
pullers, pushers, and neutral swimmers~\cite{lauga09}. In the
following, we investigate the solvent velocity fields generated by
the self-phoretic Janus particles, as well as those generated by
self-phoretic microdimers, and in both cases the analytical
predictions are compared with simulation results. The velocity field
around a self-propelled particle can be analytically calculated from
the Navier-Stokes equation. Here, we solve the Stokes equation,
which neglects the effect of inertia due to very small Reynolds
number, and consider the incompressible fluid
condition~\cite{anderson,yang13sm}. Note that although MPC has the
equation of state of an ideal gas, the compressibility effects of
the associated flow fields have shown to be very small in the case
of thermophoretic particles~\cite{yang13sm}.  We also implicitly
assume that the standard boundary layer approximation is valid, this
is that the particle-solvent interactions are short-ranged. In order
to solve the Stokes equation, three hydrodynamic boundary conditions
need to be determined. In the particle reference frame the normal
component of the flow field at the particle surface vanishes.
Considering sufficiently large systems, it is reasonable to assume
vanishing velocity field at infinity. Finally, the integral of the
stress tensor over the particle surface has to be identified in each
geometry.

\subsection{Self-phoretic Janus particle}

For a self-phoretic Janus particle, the propulsion force balances
with the friction force due to the particle motion, such that the
integral of stress tensor over the particle surface vanishes. The
resulting velocity field reads
\begin{equation}
\v(\r)=\frac{\sigma^3}{2r^3}\left(3\frac{\r\r}{r^2}-{\bf
I}\right)\cdot\v_p,\label{field_j}
\end{equation}
with ${\bf I}$ the unit tensor, $\r$ the distance to the colloid
center, and $r=|\r|$. Note that the boundary conditions are the same
as in the case of a thermophoretic particle moving in an external
temperature gradient and therefore also the velocity
field~\cite{yang13sm}. Equation~(\ref{field_j}) indicates that the
velocity field is a source dipole, which decays fast with the
distance as $1/r^3$. It is therefore to be expected that in
suspensions of the self-phoretic Janus particles, the hydrodynamic
interactions are negligible in comparison to the effects of
concentration or temperature gradients.

\begin{figure}[h]
\centering
\begin{minipage}[c]{0.5\textwidth}
\centering
\includegraphics[width=2.3in]{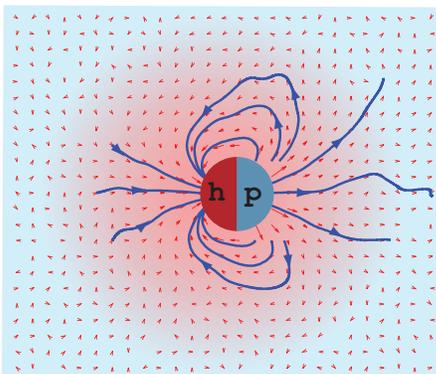}
\end{minipage}
\caption{Velocity field induced by a self-thermophoretic Janus
particle with a thermophobic surface. The left (h) and the right (p)
hemisphere corresponds to the heated and the phoretic part,
respectively. Propulsion and flow field on the axis ${\bf n}$
point in the same direction. Small arrows represents the flow
velocity magnitude and direction, and lines refer to the streamlines
of the flow field. The background color code does not precisely
correspond to the temperature distribution, and should be taken as a
guide to the eye.\label{fj}}
\end{figure}

\begin{figure}[h]
\includegraphics[width=3.in]{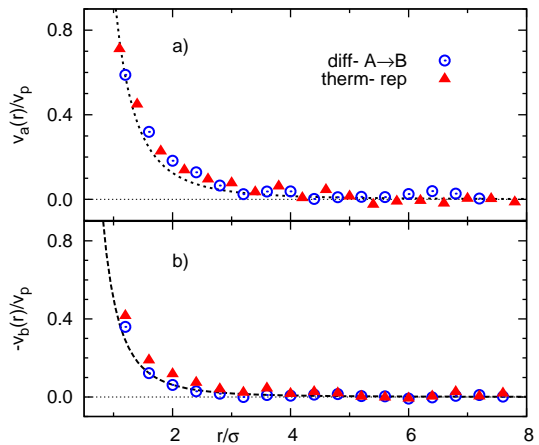}
\caption{Rescaled flow velocity, $\v\cdot\bf n$, as a function of
the distance to the center of the Janus particle (positive direction
towards the functional part). Symbols refer to the simulation
results, and lines to the predictions in Eq.~(\ref{field_j}).
a)~Velocity along the axis ${\bf n}$. b)~Velocity along the axis
perpendicular to ${\bf n}$.}\label{cj}
\end{figure}

Direct measurements of the flow field around the microswimmers can
be performed in the simulations and allow a quantitative comparison
with the analytical expression. Since only small differences are
expected between the two discussed types of phoretic swimmers, we
focus in the following on the thermophoretic microswimmers.
Figure~\ref{fj} shows the velocity field induced by a
self-thermophoretic Janus particle with a thermophobic surface in a
section across the particle center. The measured velocity field has
a source-dipole type pattern, in which propulsion and flow field
along the particle axis have the same direction as expected from the
analytical prediction in Eq.~(\ref{field_j}). The quantitative
values of the simulated velocity fields are compared with the
analytical predictions in Fig.~\ref{cj} for both the
self-thermophoretic and the self-diffusiophoretic Janus particle.
The flow field component along the Janus particle axis, $\v\cdot\bf
n$, is displayed along the Janus particle axis ${\bf n}$ in
Fig.~\ref{cj}~a and perpendicular to it in Fig.~\ref{cj}~b.
Simulation results and analytical predictions are in very good
agreement without any adjustable parameter, although on the axis
perpendicular to ${\bf n}$ the theory slightly underestimates the
velocity field of the self-thermophoretic Janus particle at short
distances. The underestimation probably arises from the sharp change
of the solvent properties at the border between the functional and
non-functional hemispheres~\cite{wurger13pre}, which is disregarded
in the present analytical calculation.

\subsection{Self-phoretic microdimer}

Besides the Janus particle, other particle geometries have been
shown to be easy to construct phoretic swimmers. Such an alternative
is the microdimer~\cite{kap07b,yang11}, composed of two strongly
attached beads, in which one bead acts as the functional end, and
the other bead as the non-functional one. The Stokes equation can be
solved independently for each bead, and the total velocity field
around the self-propelled microdimer can be approximated as a
superposition of these two velocity fields. The dimer is a typical
force dipole such that integral of stress tensor over each bead is
non-zero, although their sum vanishes. This is fundamentally
different from the case of the Janus
particle~\cite{golestanian12,wurger13pre}. The integral over the
functional bead corresponds to the frictional force, which is
associated with the propulsion velocity by $\gamma v_{p}$, with
$\gamma_p$ the friction coefficient. The integral over the
non-functional bead corresponds to the driving force which has the
same magnitude as the friction force, but opposite direction; this
results in zero net force on the dimer. By solving the Stokes
equation, the velocity field produced by the functional and
non-functional beads are
\begin{equation}
\v_{f}(\r)=\frac{\sigma}{2\mid\r-\r_{f}\mid}\left(\frac{(\r-\r_{f})
(\r-\r_{f})}{\mid\r-\r_{f}\mid^2}+{\bf
I}\right)\cdot\v_p,\label{field_da}
\end{equation}
and
\begin{eqnarray}\label{field_dp}
\v_{nf}(\r)=-\frac{\sigma}{2\mid\r-\r_{nf}\mid}\left(\frac{(\r-\r_{nf})
(\r-\r_{nf})}{\mid\r-\r_{nf}\mid^2}+{\bf I}\right)\cdot\v_p\nonumber\\
+\frac{\sigma^3}{\mid\r-\r_{nf}\mid^3}
\left(\frac{3(\r-\r_{nf})(\r-\r_{nf})}{\mid\r-\r_{nf}\mid^2}-{\bf
I}\right)\cdot\v_p,\nonumber\\
\end{eqnarray}
respectively. Here, $\r_{f}$ and $\r_{nf}$ are the position coordinates of the
functional and non-functional beads, respectively. Note that the
second term on the right side of Eq.~(\ref{field_dp}) corresponds to a
source dipole, which arises from the excluded volume effect of the
bead (vanishing for point particle). Thus, the total velocity field
around the self-propelled microdimer can be approximated by
\begin{equation}
\v(\r)=\v_{f}(\r)+\v_{nf}(\r) \sim 1/r^2.\label{field_dsum}
\end{equation}
Consequently, in suspensions composed of phoretic microdimers the
hydrodynamic interactions are comparable to the contributions coming
from concentration or temperature gradients. Furthermore, the
near-field hydrodynamic behaviors of the dimer also differ remarkably
from the Janus particle.

\begin{figure}[h]
\includegraphics[width=2.3in]{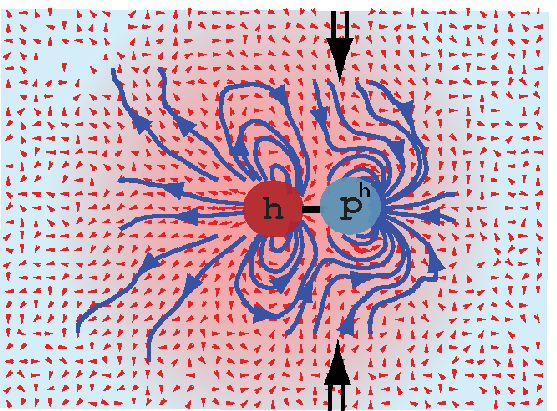}
\includegraphics[width=2.3in]{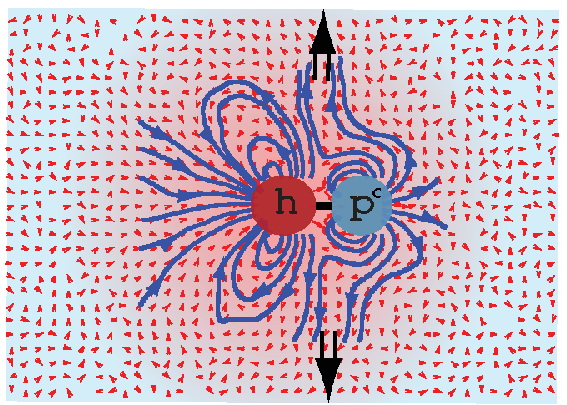}
\caption{Solvent velocity field and stream lines induced by
self-thermophoretic microdimers. a) Pusher-type of swimmer for a
thermophilic microdimer. b) Puller-type of swimmer for a
thermophobic microdimer. The left bead (h) corresponds to the heated
bead, and the right (p) to the phoretic or propelling one, $p^h$
stands for thermophilic bead, and $p^c$ for
thermophobic.}\label{fd1}
\end{figure}

Simulations of self-thermophoretic dimers allow us to perform
precise measurements of the induced flow field. The simulation model
is the same one as employed in our previous work~\cite{yang11} where
each bead has a radius $\sigma=2.5a$ and the distance between the
beads centers is $d=|\r_{nf}-\r_{f}|=5.5a$. The interactions between
the beads and the solvent are of Lennard-Jones type, cf.
Eq.~(\ref{lj}). The heated bead interacts with the solvent through a
repulsive potential ($C=\epsilon$ and $k=24$), while for the
phoretic bead two different interactions have been chosen, an
attractive ($C=0$ and $k=48$) and a repulsive interaction
($C=\epsilon$ and $k=3$). The solvent velocity field is computed
around the dimer and displayed for dimers with both interaction
types in Fig.~\ref{fd1}. In spite of the opposite orientations and
the difference in intensity, the pattern of the two flow fields are
very similar. The velocity field on the axis across the dimer center
and perpendicular to the symmetry axis is, for the microdimer with
thermophilic interactions (repulsive), oriented towards the dimer
center, while for the thermophobic dimer (attractive interactions)
is oriented against the dimer center. This is consistent with the
well-known hydrodynamic character of force dipoles~\cite{lauga09},
and has further important consequences. If another dimer or particle
is placed lateral and close to the dimer, the flow field will exert
certain attraction in the case of a thermophilic microdimer and
certain repulsion in the case of a thermophobic microdimer, which
allows us to identify them respectively as {\em pushers} and {\em
pullers}.

\begin{figure}[h]
\centering
\begin{minipage}[c]{0.5\textwidth}
\centering
\includegraphics[angle=0,width=3.3in]{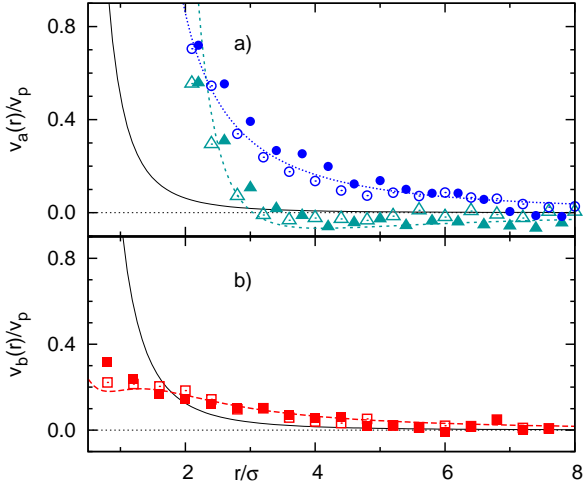}
\end{minipage}
\caption{Rescaled flow velocity as a function of distance to the
dimer center of mass. Symbols refer to the simulation results, and
discontinuous lines to the theoretical prediction in
Eq.~(\ref{field_dsum}). Solid symbols regard dimers with a
thermophilic bead and a pusher-like behavior. Open symbols regard
dimers with a thermophobic bead and a puller-like behavior. For
comparison, thin solid lines corresponds to the flow of the Janus
particle in Eq.~(\ref{field_j}). a)~Velocity along the dimer axis.
Triangles and circles correspond to the velocities on the left and
right sides of the dimer center, respectively. b)~Velocity
perpendicular to the dimer axis, with positive direction pointing to
the dimer center. }\label{cd}
\end{figure}

A quantitative comparison of the simulated velocity fields with the
analytical prediction in Eq.~(\ref{field_dsum}) is presented in
Fig.~\ref{cd} for both a pusher- and puller-type microdimer. The
flow field component on the microdimer axis analyzed along such axis
is displayed in Fig.~\ref{cd}a for the left and right branches.  The
flow field component perpendicular to the microdimer axis analyzed
along such axis is displayed in Fig.~\ref{cd}b. The small deviations
along the perpendicular axis are mostly due to statistical errors in
the simulations, while those on the bond-axis are slightly larger.
This can be attributed to the fact that the superposition
approximation in Eq.~(\ref{field_dsum}) is less precise in the case
of nearby beads. In spite this consideration, Fig.~\ref{cd} shows in
all cases that the analytical solution of the Stokes equation agrees
very nicely with the results from the MPC simulations without any
adjustable parameter, which constitutes a convincing validation for
both the analytical approximations and the model employed in the
simulations.

\section{Conclusions}

A coarse grained model to simulate a synthetic self-phoretic Janus
particle in which hydrodynamic interactions are consistently
implemented is here proposed and analyzed. The Janus particle is
provided with a proper rotation dynamics through stick particle
boundary conditions. These are modeled by bounce-back collisions
which reverse the direction of motion of the solvent particle with
respect to the moving colloidal surface. The collisions are imposed
to conserve linear and angular momentum, as well as kinetic energy.
A strong self-phoretic effect is realized by using a soft
particle-solvent potential implemented in a larger interaction
distance than the bounce-back collisions. With this model both the
self-thermophoretic and the self-diffusiophoretic Janus particles
are simulated in an straightforward manner, which further justifies
the model validity. Simulations to quantify the flow fields induced
by the self-phoretic Janus and dimer microswimmers are then also
performed, and satisfactorily compared with corresponding analytical
predictions. The flow field around the self-phoretic Janus particle
shows to be short ranged, as it is typical from neutral swimmers. In
contrast, self-phoretic microdimers induce a long-ranged flow field.
Dimers propelled towards the functional bead, as thermophilic
microdimers, show a hydrodynamic lateral attraction typical from
pushers. Conversely, dimers propelled against the functional bead,
as thermophobic microdimers, show a hydrodynamic lateral repulsion
typical from pullers. These fundamental differences will result in
systems with very different collective properties, for which our
simulation model is very adequately suited.

\section*{Acknowledgments}
M.Y. acknowledges partial support from the ``100 talent plan'' of
Institute of Physics, Chinese Academy of Sciences, China.

\appendix

\section{Bounce-back with a moving spherical particle}
\label{appendix}
Considering the contact velocity in Eq.~(\ref{relativ.vel}) and the
post-collision quantities in Eq.~(\ref{bounce}), the post-collision
contact velocity can be calculated as
\begin{equation}
\widetilde{\v}'= \widetilde{\v} - \frac{\bf p}{\mu}
+ \frac{1}{\chi M} \big[\hat{\bf s}
(\hat{\bf s}\cdot {\bf p}) - {\bf p}\big],
\end{equation}
where the relation of the vector triple product with the scalar
product has been employed. The difference between the relative pre-
and post-collision velocity,
$\Delta\widetilde{\v}=\widetilde{\v}'-\widetilde{\v}$, can be
decomposed into a normal and a tangential component as
\begin{align}
\label{ap:vn}
\Delta\widetilde{\v}_n =& -\frac{1}{\mu} \hat{\bf s}(\hat{\bf s}\cdot{\bf p}) \\
\Delta\widetilde{\v}_t =& \big[\hat{\bf s}(\hat{\bf s}\cdot{\bf p}) -{\bf p} \big]
\left(\frac{1}{\mu} + \frac{1}{\chi M} \right),
\end{align}
with which ${\bf p}$ can expressed as
\begin{equation}
\label{ap:p}
{\bf p}=\mu\left(\Delta\widetilde{\v}_n
+\frac{\chi M}{\chi M+\mu}\Delta\widetilde{\v}_t\right),
\end{equation}

The difference in kinetic energy before and after the collision can be
calculated from the pre- and post-collision velocities in Eq.~(\ref{bounce})
as
\begin{equation}
\Delta E= - 2 {\bf p}\cdot\widetilde{\v} + \frac{{\bf p}^2}{\mu} +
\frac{1}{\chi M} \left[ {\bf p}^2 - (\hat{\bf s}\cdot{\bf p})^2\right],
\end{equation}
where the circular shift property of the mixed product has been used.
Employing the expression of $\Delta\widetilde{\v}_n$ in
Eq.~(\ref{ap:vn}) and of ${\bf p}$ and ${\bf p}^2$ which can be
obtained from Eq.~(\ref{ap:p}), the previous expression can be rewritten as
\begin{align}
\Delta E= &\  \frac{\mu}{2}\left(2\widetilde{\v}+\Delta\widetilde{\v}_n\right)\cdot\Delta\widetilde{\v}_n\nonumber\\
&  +\frac{1}{2}\frac{\chi M}{\chi
M+\mu}\left(2\widetilde{\v}+\Delta\widetilde{\v}_t\right)\cdot\Delta\widetilde{\v}_t.
\end{align}
To ensure a collision with energy conservation, it is necessary that
both components of the previous expression vanish, since the
prefactors are determined by the system under study. Using
orthogonality of normal and tangential velocity components the two
previous conditions translate into,
$\widetilde{\v}_n^2=\widetilde{\v}_n^{\prime 2}$ and
$\widetilde{\v}_t^2=\widetilde{\v}_t^{\prime 2}$.  Two physical
meaningful solutions exist, both with
$\widetilde{\v}_n=-\widetilde{\v}_n^\prime$. One is the specular
reflection of smooth hard spheres,
$\widetilde{\v}_t=\widetilde{\v}_t^\prime$, which is well-known to
imply slip-boundary condition. Another solution is the bounce-back
reflection of rough hard spheres,
$\widetilde{\v}_t=-\widetilde{\v}_t^\prime$, which enforces a
no-slip boundary condition between the solvent and the solute. With
both conditions it is possible to express $\Delta\widetilde{\v}$ and
hence ${\bf p}$ in terms of the components of the pre-collision
contact velocity $\widetilde{\v}$, which is specified in
Eq.~(\ref{momentum}) for the no-slip condition employed in this
work.


\end{document}